\documentclass{vgtc}                         

\ifpdf
  \pdfoutput=1\relax                   
  \usepackage{graphicx}                
  \DeclareGraphicsExtensions{.pdf,.png,.jpg,.jpeg} 
\else
  \usepackage{graphicx}                
  \DeclareGraphicsExtensions{.eps}     
\fi%

\graphicspath{{figures/}{pictures/}{images/}{./}} 

\usepackage{microtype}                 
\PassOptionsToPackage{warn}{textcomp}  
\usepackage{textcomp}                  
\usepackage{mathptmx}                  
\usepackage{times}                     
\usepackage{cite}                      
\usepackage{tabu}                      
\usepackage{booktabs}                  

\usepackage{wrapfig}
\usepackage{enumitem}
\usepackage{tabularx}
\usepackage{diagbox}
\usepackage[symbol]{footmisc}

\usepackage{xcolor}

\newcommand{\familiarity}[0]{{\textsc{familiarity}}}
\newcommand{\clarity}[0]{{\textsc{clarity}}}
\newcommand{\credibility}[0]{{\textsc{credibility}}}
\newcommand{\reliability}[0]{{\textsc{reliability}}}
\newcommand{\confidence}[0]{{\textsc{confidence}}}

\onlineid{1066}

\vgtccategory{Research}

\vgtcinsertpkg

\title{Do You Trust What You See? \\Toward A Multidimensional Measure of Trust in Visualization}

\vspace{-4mm}
\author{
Saugat Pandey, Oen G. McKinley \thanks{contributed equally}~\thanks{ e-mail: p.saugat@wustl.edu, m.oen@wustl.edu}\\ %
     \scriptsize Washington University in St. Louis %
\and R. Jordan Crouser\thanks{e-mail: jcrouser@smith.edu}\\ %
        \scriptsize  Smith College %
\and Alvitta Ottley\thanks{e-mail: alvitta@wustl.edu}\\ %
     \scriptsize Washington University in St. Louis %
}

\teaser{
  \centering
  \vspace{-3mm}
  \includegraphics[width=.9\linewidth]{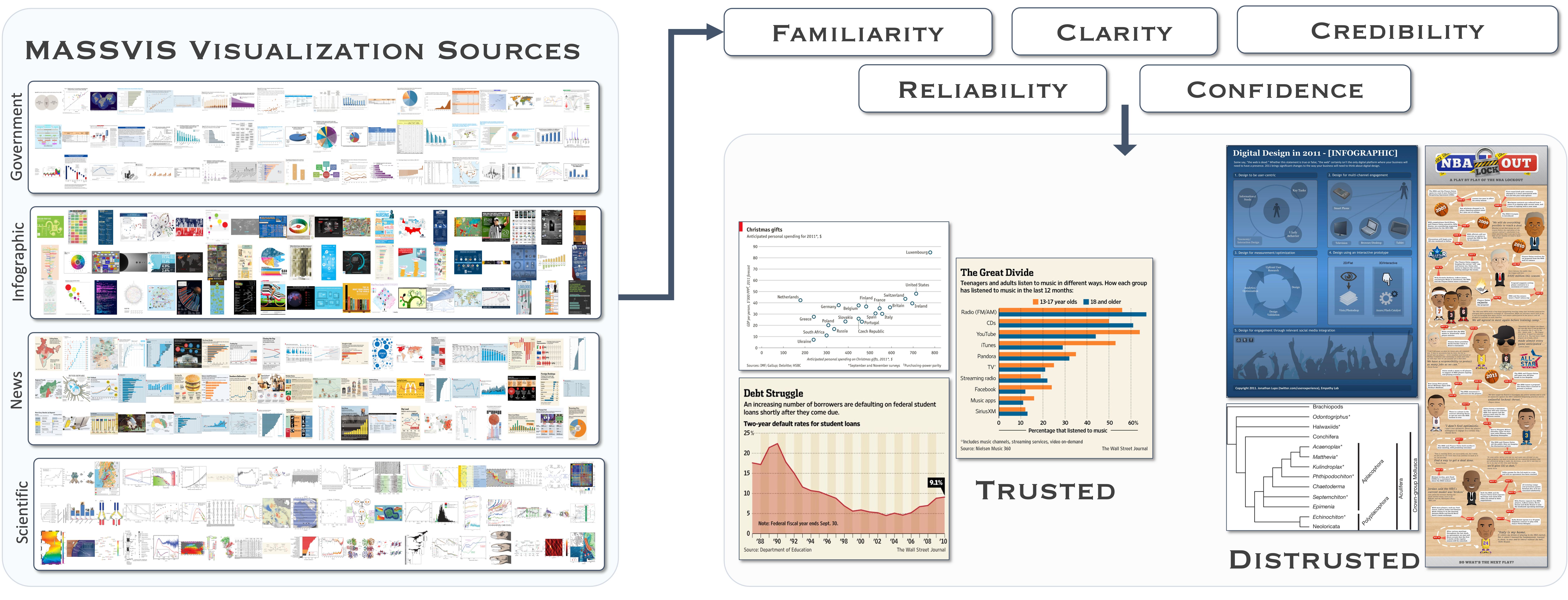}
  \vspace{-4mm}
  \caption{An illustrative overview of our experiments. Participants rated a diverse set of visualizations based on their \familiarity, \clarity, \credibility, \reliability, and \confidence. We examined how these ratings align with visual features and trust rankings.}
  \label{fig:teaser}
}

\abstract{

Few concepts are as ubiquitous in computational fields as trust. However, in the case of information visualization, there are several unique and complex challenges, chief among them: defining and measuring trust. In this paper, we investigate the factors that influence trust in visualizations. We draw on the literature to identify five factors likely to affect trust: credibility, clarity, reliability, familiarity, and confidence. We then conduct two studies investigating these factors' relationship with visualization design features. In the first study, participants' credibility, understanding, and reliability ratings depended on the visualization design and its source. In the second study, we find these factors also align with subjective trust rankings. Our findings suggest that these five factors are important considerations for the design of trustworthy visualizations. 
} 

\CCScatlist{
  \CCScatTwelve{Human-centered computing}{Visu\-al\-iza\-tion}{Visualization design and evaluation methods}{}
}

\begin{document}

\firstsection{Introduction}

\maketitle

Trust is an amorphous concept but is also an integral part of human interactions with machines. Many fields in Computer Science and Engineering have explored the topic of trust in recent decades, including Machine Learning \cite{ML}, Automation and Robotics \cite{Auto}, and Information and Communication Technology \cite{ICT}. Research in information visualizations has recently adopted a similar focus, with work aimed at finding the context where trust can be best applied to the field \cite{MHSW20, HS20}, often pulling from other disciplines such as psychology \cite{BE20} or related fields such as cartography \cite{G23}. 

These prior works highlight the need to better understand the role of trust in visualization. They suggest that the effectiveness of visualization depends on the user's trust in the data and representation. When users perceive a visualization as untrustworthy, they may hesitate to rely on the information presented or take action based on it. This could lead to poor decision-making, missed opportunities, or dangerous outcomes for high-stakes decisions.  However, little research is on the factors influencing users' trust or guidance for designing more trustworthy visualizations.  Further, there is currently no standard approach for defining and evaluating the trustworthiness of visualizations, making it difficult to compare and evaluate different visualizations consistently and objectively~\cite{BE20, HS20, MHSW20}.
This paper takes a step toward defining and evaluating trust in information visualization, drawing upon prior work in various fields (e.g., AI, human-robot interaction, and psychology). We focus on the \textit{trust perception} and behavior of users instead of the ``objective" trustworthiness of the data or visualization. As a result, we expect that the visualization design will likely substantially affect perceived trust. 

We present the findings of two user studies. 
Experiment~1 (Exp.~1) examined the correlation between visualization design features and five trust dimensions: credibility, clarity, reliability, familiarity, and confidence. Excluding familiarity, the trust dimensions correlated with the visualization's design features and source. 
Experiment~2 (Exp.~2) was a follow-up study to understand how people describe trust in the context of visualization design. We asked participants to rank a collection of visualizations based on how much they trusted them and then explain their ranking. We then coded their responses and found that they often related to the five trust dimensions from Exp. 1. 
For example, the comments aligned with the credibility dimensions discussed the importance of visualizations labeled with the data source. Some participants talked about the importance of visualizations being easy to understand, mirroring clarity. These findings suggest a positive relationship between the five trust dimensions and perceived trust in visualizations. We discuss how these results may help designers improve trust in their visualizations.

\section{Approaches to defining and measuring trust}

Trust has received a great deal of attention, with researchers proposing many definitions and conceptualizations of trust~\cite{lewicki2006models}. Although it is impossible to capture every existing definition of trust, we highlight some key perspectives.

\subsection{A social sciences perspective}
The Oxford English Dictionary defines trust as ``firm belief in the \textbf{reliability}, truth, ability, or strength of someone or something.'' Alternatively, personality psychologist Rotter\cite{rotter1967new}  defines trust as a personality trait that describes one's ``generalized expectancy that the oral or written statement of other people can be \textbf{relied} on.''\cite{rotter1967new} Both highlight the common expectation of \textbf{reliability} as an integral component of trust. In contrast, sociologists view trust as a social construct, classifying it as a group property rather than an individual one~\cite{lewis1985trust}. Through this lens, Luhmann~\cite{luhmann2018trust} argues that ``\textbf{familiarity} is the precondition for trust and distrust,'' reasoning that one must be familiar with the object of trust. Similarly, Simmel~\cite{simmel2004philosophy} asserts that when faced with something unfamiliar, we ``gamble'' rather than ``trust''. Simmel aligns with behavioral economists' view of trust as ``the mutual \textbf{confidence} that no party to an exchange will exploit another’s vulnerabilities''~\cite{sabel1993studied}. This notion of confidence, particularly in decision-making under uncertainty, appears in several of the prior works~\cite{barney1994trustworthiness,delgado2003development,garbarino1999different,crosby1990relationship}. Another example from Morgan and Hunt~\cite{morgan1994commitment} defines trust as the perception of ``\textbf{confidence} in the exchange partner's \textbf{reliability} and \textbf{integrity}.''  

Researchers have proposed assessing trust as an enduring trait in people and studying trust as a broad construct that complements personality models such as the Big Five Inventory \cite{rousseau1998not, digman1997higher}. For instance, Evans and Revelle created and tested a 21-item scale that measures interpersonal trust, with items such as ``Listen to my conscience'' and ``Avoid contact with others'' \cite{evans2008survey}. Further, research in organizational psychology proposed several dimensions of trust, including \textbf{integrity}, \textbf{reliability}, and \textbf{credibility},  based on a comprehensive review of existing literature on trust measurement \cite{mcevily2011measuring}.

\subsection{An AI and Human-Robot Interaction Perspective}
We also considered work from the Artificial Intelligence (AI) and Human-Robot Interaction (HRI) communities to inform how we measure trust in information visualization. Like with visualization, the effectiveness of incorporating AI into organizations is significantly dependent on users' trust in the technology \cite{hoffman2013trust}.

In AI, trust can be defined as the willingness of an individual to rely on and be vulnerable to the actions of another party based on the expectation that the other will perform a particular action that is important to the trustor, irrespective of the ability to monitor or control that other party \cite{glikson2020human, mayer1995integrative}. 
Researchers measure trust in AI through various means, including user reactions to technological features, perceived usefulness, ease of use, and trust as a predictor of technology acceptance \cite{ghazizadeh2012extending, pavlou2003consumer}. 
Although the antecedents of trust in AI are still not fully understood, researchers have identified several dimensions of trust that are relevant in the context of AI, including \textit{tangibility, transparency, \textbf{reliability}, task characteristics, and immediacy behaviors} \cite{glikson2020human}.

We observed more progress in the HRI community, where researchers have developed various methods for measuring trust. For example, some studies ask participants to rate their agreement to statements such as ``I trust that the robot can perform the task successfully'' \cite{rossi2018impact,shu2018human}. Others have sought to create standardized and generalizable scales for measuring trust, such as \textit{The Reliance Intention Scale}~\cite{lyons2019individual} and \textit{The Trust Perception Scale-HRI (TPS-HRI)}~\cite{schaefer2016measuring}.
Work by Ullman and Malle and their Multidimensional Measure of Trust (MDMT)~\cite{ullman2018does} is also notable. They began by surveying the cross-disciplinary perspectives of trust, summarizing definitions from dictionaries, human-automation, human-human, and human-robot interaction research. They found that dictionaries typically mention terms such as ability, \textbf{reliability}, honesty, and \textbf{integrity}, and argue that trust in HRI should capture four dimensions: \textit{Reliable, Capable, Ethical, and Sincere}.

\begin{table}[b!]
\vspace{-2mm}
\centering
\scriptsize
\caption{The proposed trust dimensions. We are informed by the overlaps in how trust is categorized in the literature.}
\label{tab:dimensions}
\begin{tabularx}{\linewidth}{lp{6cm}}
\toprule

\textsc{Familiarity:} & The extent to which users are familiar with the data, topic, and visualization. \textit{``I am familiar with the topic or data this visualization presents.''} \cite{luhmann2018trust} \\
\textsc{Clarity:} & The extent to which users can easily understand the visualization. \textit{``I understand what this visualization is trying to tell me.''} \cite{mayr2019trust} \\
\textsc{Credibility:} & The extent to which users believe the visualization is accurate and presents real data. \textit{``I believe the visualization shows real data.''} \cite{morgan1994commitment, ullman2018does, mcevily2011measuring} \\
\textsc{Reliability:} & The extent to which users believe the visualization will support sound decisions. \textit{``I would rely on the facts in this Visualization.''} \cite{ rotter1967new,lewis1985trust,morgan1994commitment,mcevily2011measuring,glikson2020human,ullman2018does,mayr2019trust} \\
 \textsc{Confidence:} & The extent to which users feel confident interpreting the visualization. \textit{``I would feel confident using the information to make a decision.''} \cite{sabel1993studied, morgan1994commitment} \\
\bottomrule
\end{tabularx}
\end{table}

\subsection{Trust in Visualization}

Trust is key in determining whether a user will rely on and build on the information provided in a visualization. Researchers of visualizations acknowledge the complexities of trust and its evaluation, and they employ various metrics to measure trust \cite{elhamdadi2022we, sacha2015role, xiong2019examining, zhou2019effects}. This ambiguity in measuring trust makes it difficult to assess the validity of these different measures and whether they capture trust. Without a shared understanding of what trust is and how it can be measured, it becomes challenging to evaluate trust, hindering the ability of researchers to make meaningful conclusions across studies. 

The most common approach to measuring trust is directly asking users to indicate their level of trust using a Likert scale \cite{xiong2019examining, kim2017data,zhou2019effects,padilla2022multiple}. Kim et al. \cite{kim2017data} used a 100-point scale to measure participants' trust in the accuracy of the data. Similarly, Zhou et al. \cite{zhou2019effects} used a 9-point scale to measure the users' trust towards predictive decision-making. In addition to direct measures of trust, such as asking users to rate their level of trust in a visualization, researchers have also used proxy measures, such as substitution variables. For instance, Xiong et al. \cite{xiong2019examining} presented participants with various map visualizations for a hypothetical firetruck emergency and asked them to choose the most trustworthy one. They also collected ratings of perceived trust and transparency, which encompassed four key dimensions: the accuracy of the information, the clarity of the communication, the amount of information disclosed, and the extent to which the shared information reflected the underlying data. These methods allow for a more nuanced understanding of how users perceive the trustworthiness of visualizations. 

Moreover, Mayr et al. \cite{mayr2019trust} provided a definition of trust that emphasized the importance of \textbf{reliability} and \textbf{understanding} for visualizations. They characterize trust as the users' tendency to rely on and build on the information displayed in a visualization. This concept emphasizes that a user's trust is intimately related to their impression of the veracity and usefulness of the information supplied. Therefore, visualizations that are perceived as reliable and easy to understand are more likely to be trusted, while those that are perceived as deceptive and complicated are less likely to be trusted. These considerations highlight the importance of developing an effective approach to measuring trust in visualizations.

\begin{figure*}[h!]
\centering
\includegraphics[width=\textwidth]{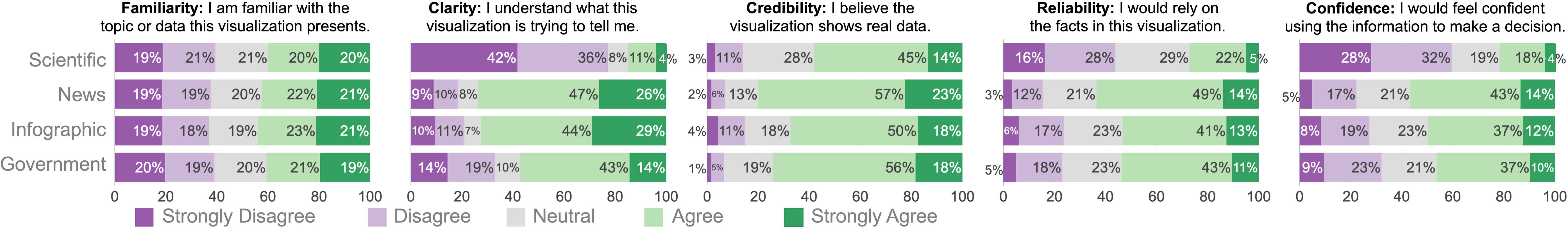}
\vspace{-6mm}
\caption{A comparison of participants' responses across four source categories (scientific, news, infographics, and government) for five dimensions in data visualization. We can observe that the \familiarity\ rating was similar across all visualization categories. Additionally, visualizations from news media were generally rated as clear, credible, reliable, and elicited high confidence levels.}
\label{fig:correlation}
\vspace{-3mm}
\end{figure*}

\section{A Multidimensional Measure of Trust in \\Visualization}
\label{dimension}

Most conceptualizations of trust endorse a multidimensional notion~\cite{mcevily2011measuring,ullman2018does}. 
Also, many fields reason about the relationship between trustors and trustees~\cite{glikson2020human, mayer1995integrative,hoffman2013trust}, which does not directly translate to scenarios where the visualization's source is ambiguous or unknown. Users may trust a visualization regardless of who created it if it is transparent, easy to understand, and based on credible data. Conversely, users may distrust a visualization even if it comes from a trustworthy source, if it is poorly designed or based on unreliable data. Therefore, we aim to develop objective criteria for evaluating visualizations independent of the biases toward the source or creator.
Table~\ref{tab:dimensions} details five critical dimensions of trust identified in the literature: credibility, clarity, reliability, familiarity, and confidence.

\section{Methods}

We present the findings of two user studies.
Exp. 1 examines the correlation between the five trust dimensions and visualization design features. Exp. 2 aims to understand how people describe trust in the context of visualization design and how well these descriptions align with the five dimensions. Together, they provide a more comprehensive understanding of trust in information visualization and inform the development of more trustworthy and effective visualizations.

\subsection{Experiment 1: Measuring the Relationship between Trust Dimensions and Visualization Design}
 To examine the relationship between the visualization designs and trust dimensions, we recruited 500 participants from Prolific.
 They were all English-speaking from the USA and received \$5 for completing the study.
 255 self-identified as males and 240 as females, with a self-reported age range of 18 to 77. We used 410 visualizations from the MASSVIS dataset \cite{borkin2013makes}. We selected this dataset because the visualizations were collected from a diverse set of sources (e.g., government, news venues, scientific publications), and the visualizations were pre-coded with labels indicating features such as whether they included human-recognizable objects and the data-ink ratio.

{\textbf{Procedure and Design}} \hspace{0.3cm} The experiment was divided into two sections: (1) Asking participants to label the visualizations and (2) Measuring visualization literacy using the Mini-VLAT. The first section asks participants to label 30 visualizations randomly selected from the corpus of 410 visualizations using a 5-point Likert scale based on statements mentioned in Table~\ref{tab:dimensions}. An attention check was included to ensure engagement. In the second section, participants performed the Mini-VLAT \cite{pandey2023mini} to measure their visualization literacy and explore potential relationships between their responses and different dimensions.

Participants had unlimited time to complete the study, spending an average of 77.3 seconds ($\sigma = 98.1s$) viewing each visualization, with a minimum time of 10.1 seconds. Using Pearson's $r$ coefficients, the correlation test revealed no statistically significant correlation between the time taken to view a visualization and the participants' Likert scale ratings for each dimension ($p > 0.1$ in all cases). Therefore, the duration of participants' viewing did not impact their ratings or perception of different visualization dimensions. 



{\textbf{Data}} \hspace{0.3cm} The variables for this experiment are: 

\vspace{-.75em}
\begin{itemize}[noitemsep]
    \item \textbf{trust dimension}: \{ \familiarity, \clarity, \credibility, \reliability, \confidence \} $\in [1...5]$
    \item \textbf{visualization literacy} $\in [0...12]$
    \item \textbf{categories}: \{ scientific, government, news, infographics \}

    \item \textbf{attributes} Black \& White $\in [yes, no]$; Number of Distinct Colors $\in [1, 2-6, \geq 7]$; Data-Ink Ratio $\in [good, medium, bad]$; Visual Density $\in [low, medium, high]$; Human Recognizable Objects $\in [yes, no]$; Human Depiction $\in [yes, no]$.
\end{itemize}

{\textbf{Results}} \hspace{0.3cm} We began by examining the relationship between the visualization categories and participants' subjective ratings based on the five trust dimensions.  We ran Kruskal-Wallis tests for each dimension to examine differences in subjective ratings according to the visualization source. Except for \familiarity, there was a significant relationship between the visualization's source and the dimensions, all with $p < .001$. Post-hoc Mann-Whitney tests using a Bonferroni adjusted alpha ($\alpha = .0083$) uncovered noteworthy patterns. For example, \textit{scientific} visualizations yielded significantly lower ratings than all other categories in \clarity, \credibility, \reliability, and \confidence. We observed similar patterns with visualizations from \textit{government} agencies, especially for \clarity. Conversely, visualizations from \textit{news} media elicited significantly higher \credibility, \reliability, and \confidence\ scores than all other categories.  This trend can be seen in \autoref{fig:correlation}, and we provide more analysis details in our supplementary materials.

We then examined the relationship between visual features and trust dimensions. Again, we conducted Kruskal-Wallis and post hoc Mann-Whitney tests for each dimension with the appropriate Bonferroni-adjusted alphas. When examining the \textit{\# of distinct colors} and \textit{data-ink ratio}, we found no significant association with \familiarity, \credibility, \reliability, and \confidence. However, we observed a significant difference in \clarity rating \textit{\# of distinct colors} and \textit{data-ink ratio} attributes. Similarly, \textit{black \& white} visualizations received significantly lower ratings for \clarity, \reliability, and \confidence\ compared to colorful ones. In general, colorful and higher data-ink ratio visualizations were favored.

Finally, we observed a low positive correlation (\textit{r = 0.31}, \textit{p $<$ 0.001}) between the participants' ratings on \clarity and their visualization literacy scores. This provides suggestive evidence that higher visualization literacy was correlated with understanding the message conveyed by the visualizations. This finding highlights the importance of considering the audience's visualization literacy when designing and evaluating visualizations.

\subsection{Experiment 2: Exploring the Relationship between Trust and Individual Dimensions}

Exp. 2 aims to understand how people describe trust in the context of visualization design. Although the dimensions from Exp. 1 were curated from a diverse set of existing literature, we also use this study to examine how well the observed rating aligns with perceived trust.
To achieve this, we selected six visualizations based on each of the dimensional values obtained from Exp. 1. Specifically, we chose the visualizations with the three highest and three lowest mean scores for the given dimension, provided that they had at least 17 respondents (top 10 percentile) and a standard deviation that was lower than the median. In other words, after filtering for the visualizations with high response rates and low variance, we chose the visualizations that were, on average, ranked to be the best and worst for the given dimension. This resulted in 30 visualizations (six for each dimension). We then examined how trust values were related to the individual dimension values for these selected visualizations. 

{\textbf{Procedure and Design}} \hspace{0.3cm} We recruited 15 participants using convenience sampling, all of whom had bachelor's degrees, with nine self-identified males and six self-identified females. We used neutral language to avoid unintentional bias; the information sheet and questions excluded any mention of the five dimensions from Exp. 1. We asked participants to ``Rank the following visualizations based on how much you trust them, with 1 being Strong Trust and 6 being Strong Distrust.'' Then we asked them to ``Please explain your ranking.'' Each participant ranked five sets of six visualizations corresponding to the trust dimensions from Exp. 1. The orders of the options and questions were completely randomized.



\begin{wrapfigure}[27]{l}{0.21\textwidth}
  \begin{center}
    \raisebox{0pt}[\dimexpr\height-1.7\baselineskip\relax]{%
    \includegraphics[width=0.24\textwidth]{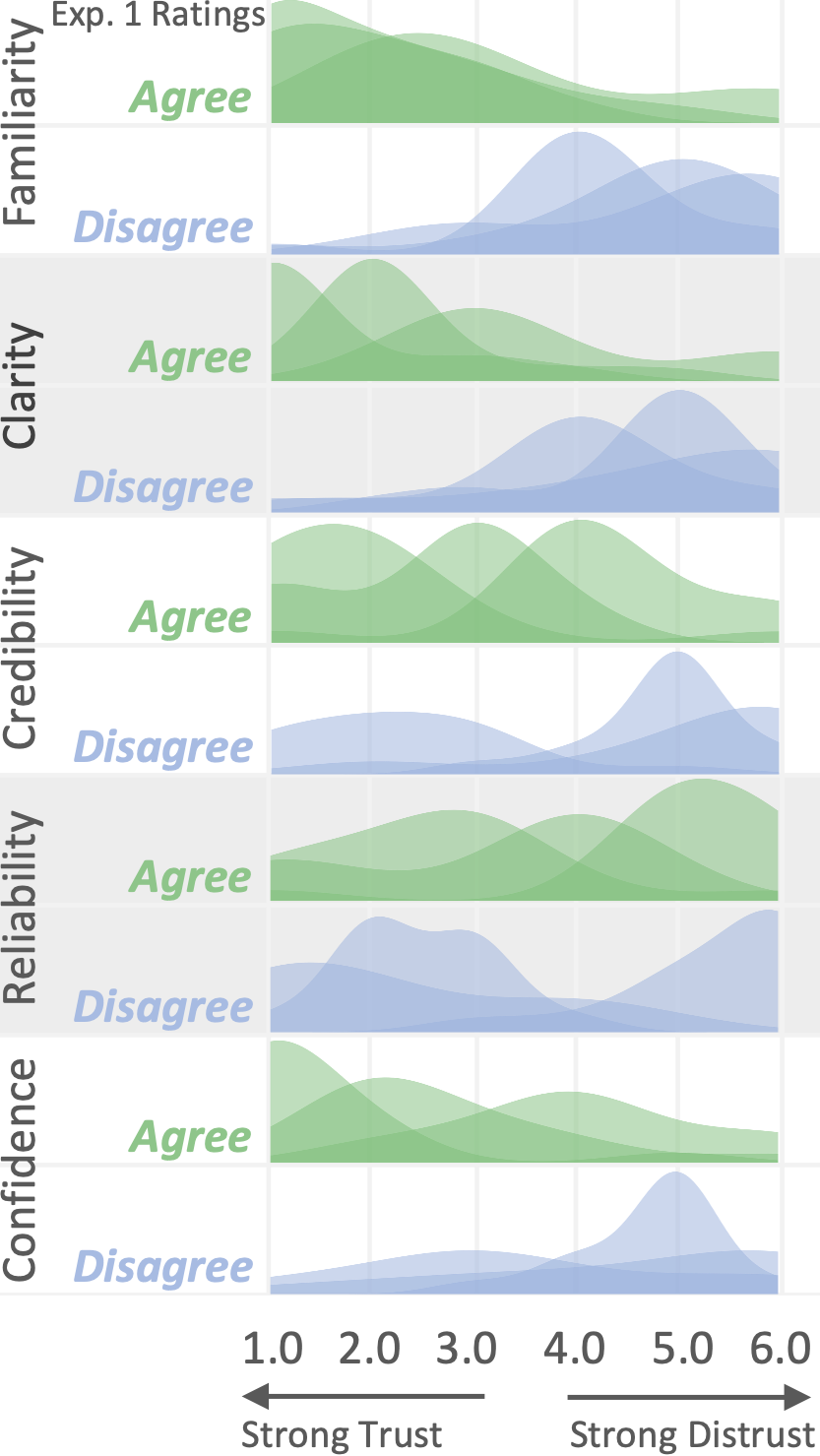}}
  \end{center}

  \begin{minipage}{.24\textwidth}
  \vspace{-5mm}
  \caption{Comparing trust ratings from Exp. 2 to the three with the highest and lowest mean scores for the dimensions in Exp. 1.}
  \label{fig:exp2}
 \end{minipage}
\end{wrapfigure}

{\textbf{Results}} \hspace{0.3cm} \autoref{fig:exp2} suggests that \familiarity\ and \clarity\ ratings from Exp. 1 strongly aligned with the perceived trust from Exp. 2, with the top three visualizations in both categories eliciting high trust ratings and the bottom three receiving low ratings. 
For \confidence\ and \credibility, two visualizations from both the \textit{agree} and \textit{disagree} categories were trusted and distrusted, respectively. However, in the case of \reliability, we observed no obvious separation in the trust ratings.
There could be several reasons we observed a different pattern for \reliability, e.g., the differences in our study populations. The concept of \textit{reliability} is complex and multifaceted, and individual differences may affect interpretation. However, our experiment design traded control of individual differences for controlling learning effects and biases.

To analyze the explanations given by the participants, we used an open-ended, qualitative, and iterative approach in identifying the factors used by the participants in ranking the images based on their perceived trust towards the visualizations. Two authors coded the responses and identified seven keywords with mutual consensus: \textit{familiarity with the topic, interpretability, reliability, source present or not, visual aesthetic, type of visualization, and accuracy}. The authors achieved an inter-rater reliability of 0.947 in the first round and finalized the keywords in the second round. Notably, 12 of the 15 participants said \textit{source} was the primary reason for trusting visualizations. For example, P12 stated, \textit{``Ranking is based on [a] combination of source provided and the nature of visualizations.''} Additionally, 10 of the 15 participants explicitly referred to their familiarity with the topic as an important factor in their trust rankings. For example, P15 stated, `\textit{`my familiarity with the topic and data source''} as the rationale for trust.  For the remaining keywords, visualization type was mentioned by 8 out of 15 participants, interpretability was pointed to by 7 out of 15, visual aesthetic by 5, reliability by 4, and accuracy by 3. Interestingly, 5 of the 15 participants considered complex and long infographics untrustworthy visualizations. For example, P10 stated \textit{``I find long images with cartoons deceiving''} and P13 said \textit{``I find long images less trustworthy''}. Additionally, participants referred to infographics as \textit{``long images''}, \textit{``images with cartoons''}, and \textit{``marketing posters''} instead of using the technical term, indicating a gap in familiarity with the type of visualization and correctly naming them. These results suggest that researchers should consider these dimensions when measuring trust in data visualizations, as they align with the participants' explanations.

\section{Discussion and Future Work}


This paper argues for a multidimensional approach to measuring trust in information visualization, recognizing that trust is influenced by various factors~\cite{mayr2019trust, rotter1967new, rotter1971generalized, rotter1980interpersonal}. 
Our first study 
showed that the source and design features of visualizations can affect trust perception. Interestingly, visualizations from news media were seen as more \textit{credible} and \textit{reliable} than those from scientific or government agencies. 
This may be because news media visualizations are often designed to be more engaging and visually appealing, using storytelling elements and presentation styles that resonate better with the public. In contrast, scientific and government visualizations are often technical and data-heavy, making them less accessible. 

Participants also favored colorful visualizations 
and visual embellishments.
Many saw infographics as approachable, despite the risk that they could oversimplify complex issues or data, leading to potential misinterpretation.
Still, collaborations between scientific or government sources, news media outlets, and designers could create accurate and engaging visualizations and bridge the gap between credibility and accessibility.
Finally, Exp. 2 provides some additional guidance, showing that \textit{source, credibility, familiarity with the topic/data, and the type of visualizations} played a significant role in their rankings for trust. 

These findings highlight the complex interplay between visualization design and trust perception. Balancing these factors is crucial to effectively communicate information, foster trust, and promote accurate understanding among diverse audiences.
Future research is needed to determine the best practices for designing visually appealing visualizations with the appropriate level of complexity.  Furthermore, future work may consider visualization used in more personal decision-making contexts (e.g., medical decisions ~\cite{bancilhon2020let,bancilhon2023combining,mosca2021does,ottley2012visually}) and situational factors like time pressure, task complexity, cognitive load, and emotional states which may impact how people perceive the trustworthiness of data visualizations. Beyond visualization literacy, future work can also consider how other individual differences~\cite{liu2020survey} such as personality, cognitive abilities, education, and cultural background impact perceived trust.

{\textbf{Limitations}} \hspace{0.3cm} Although providing initial insights, future research should address limitations, such as the relatively small size in Exp. 2. Further studies are needed to systematically isolate  visual features and factors influencing trust in visualizations.  The definition of \familiarity can be expanded to encompass various forms of prior knowledge, such as familiarity with the chart type.


\section{Conclusion}
We present two experiments that provide valuable insights into the factors that impact users' trust in data visualizations. Exp. 1 highlights the significance of the visualization source and its visual features in affecting users' subjective ratings on five dimensions. Specifically, the findings suggest that visualizations produced by news media are more trusted than those produced by scientific or government agencies. In addition, colorful and high \textit{data-ink ratio} visualizations are rated higher in terms of \clarity, \reliability, and \confidence, while \textit{black and white} visualizations are rated lower. Exp. 2 demonstrates alignment between participants' ranking of visualizations based on perceived trust and the ratings of \familiarity\ and \clarity\ from Exp. 1. The results also emphasize the importance of factors such as \textit{source, credibility, familiarity with the topics, and interpretibility} in influencing users' trust in data visualizations. As such, the study highlights the need for designers of data visualizations to carefully consider these factors to build user trust in their visualizations.

\acknowledgments{
This material is based upon work supported by the National Science Foundation under grant number 2142977 and the Laboratory for Analytic Sciences (LAS).} 


\bibliographystyle{abbrv-doi}

\bibliography{template}

\begin{thebibliography}{10}

\bibitem{Auto}
H.~Azevedo-Sa, S.~K. Jayaraman, C.~T. Esterwood, X.~J. Yang, L.~P. Robert, and
  D.~M. Tilbury.
\newblock Real-time estimation of drivers’ trust in automated driving
  systems.
\newblock {\em International Journal of Social Robotics}, 2020.

\bibitem{bancilhon2020let}
M.~Bancilhon, Z.~Liu, and A.~Ottley.
\newblock Let’s gamble: How a poor visualization can elicit risky behavior.
\newblock In {\em 2020 ieee visualization conference (vis)}, pp. 196--200.
  IEEE, 2020.

\bibitem{bancilhon2023combining}
M.~Bancilhon, A.~Wright, S.~Ha, R.~J. Crouser, and A.~Ottley.
\newblock Why combining text and visualization could improve bayesian
  reasoning: A cognitive load perspective.
\newblock In {\em Proceedings of the 2023 CHI Conference on Human Factors in
  Computing Systems}, pp. 1--15, 2023.

\bibitem{barney1994trustworthiness}
J.~B. Barney and M.~H. Hansen.
\newblock Trustworthiness as a source of competitive advantage.
\newblock {\em Strategic management journal}, 15(S1):175--190, 1994.

\bibitem{BE20}
R.~Borgo and D.~J. Edwards.
\newblock The development of visualization psychology analysis tools to account
  for trust.
\newblock {\em http://arxiv.org/abs/2009.13200}, 2020.

\bibitem{borkin2013makes}
M.~A. Borkin, A.~A. Vo, Z.~Bylinskii, P.~Isola, S.~Sunkavalli, A.~Oliva, and
  H.~Pfister.
\newblock What makes a visualization memorable?
\newblock {\em IEEE transactions on visualization and computer graphics},
  19(12):2306--2315, 2013.

\bibitem{crosby1990relationship}
L.~A. Crosby, K.~R. Evans, and D.~Cowles.
\newblock Relationship quality in services selling: an interpersonal influence
  perspective.
\newblock {\em Journal of marketing}, 54(3):68--81, 1990.

\bibitem{delgado2003development}
E.~Delgado-Ballester, J.~L. Munuera-Aleman, and M.~J. Yague-Guillen.
\newblock Development and validation of a brand trust scale.
\newblock {\em International journal of market research}, 45(1):35--54, 2003.

\bibitem{digman1997higher}
J.~M. Digman.
\newblock Higher-order factors of the big five.
\newblock {\em Journal of personality and social psychology}, 73(6):1246, 1997.

\bibitem{elhamdadi2022we}
H.~Elhamdadi, A.~Gaba, Y.-S. Kim, and C.~Xiong.
\newblock How do we measure trust in visual data communication?
\newblock In {\em 2022 IEEE Evaluation and Beyond-Methodological Approaches for
  Visualization (BELIV)}, pp. 85--92. IEEE, 2022.

\bibitem{evans2008survey}
A.~M. Evans and W.~Revelle.
\newblock Survey and behavioral measurements of interpersonal trust.
\newblock {\em Journal of Research in Personality}, 42(6):1585--1593, 2008.

\bibitem{garbarino1999different}
E.~Garbarino and M.~S. Johnson.
\newblock The different roles of satisfaction, trust, and commitment in
  customer relationships.
\newblock {\em Journal of marketing}, 63(2):70--87, 1999.

\bibitem{G23}
G.~Gartner.
\newblock Towards a research agenda for increasing trust in maps and their
  trustworthiness.
\newblock {\em Kartografija i Geoinformacije}, 21, 2023.

\bibitem{ghazizadeh2012extending}
M.~Ghazizadeh, J.~D. Lee, and L.~N. Boyle.
\newblock Extending the technology acceptance model to assess automation.
\newblock {\em Cognition, Technology \& Work}, 14:39--49, 2012.

\bibitem{glikson2020human}
E.~Glikson and A.~W. Woolley.
\newblock Human trust in artificial intelligence: Review of empirical research.
\newblock {\em Academy of Management Annals}, 14(2):627--660, 2020.

\bibitem{HS20}
W.~Han and H.-J. Schulz.
\newblock Beyond trust building — calibrating trust in visual analytics.
\newblock {\em 2020 IEEE Workshop on TRust and EXpertise in Visual Analytics
  (TREX)}, p. 9–15, 2020.

\bibitem{hoffman2013trust}
R.~R. Hoffman, M.~Johnson, J.~M. Bradshaw, and A.~Underbrink.
\newblock Trust in automation.
\newblock {\em IEEE Intelligent Systems}, 28(1):84--88, 2013.

\bibitem{kim2017data}
Y.-S. Kim, K.~Reinecke, and J.~Hullman.
\newblock Data through others' eyes: The impact of visualizing others'
  expectations on visualization interpretation.
\newblock {\em IEEE transactions on visualization and computer graphics},
  24(1):760--769, 2017.

\bibitem{lewicki2006models}
R.~J. Lewicki, E.~C. Tomlinson, and N.~Gillespie.
\newblock Models of interpersonal trust development: Theoretical approaches,
  empirical evidence, and future directions.
\newblock {\em Journal of management}, 32(6):991--1022, 2006.

\bibitem{lewis1985trust}
J.~D. Lewis and A.~Weigert.
\newblock Trust as a social reality.
\newblock {\em Social forces}, 63(4):967--985, 1985.

\bibitem{liu2020survey}
Z.~Liu, R.~J. Crouser, and A.~Ottley.
\newblock Survey on individual differences in visualization.
\newblock In {\em Computer Graphics Forum}, vol.~39, pp. 693--712. Wiley Online
  Library, 2020.

\bibitem{luhmann2018trust}
N.~Luhmann.
\newblock {\em Trust and power}.
\newblock John Wiley \& Sons, 2018.

\bibitem{lyons2019individual}
J.~B. Lyons and S.~Y. Guznov.
\newblock Individual differences in human--machine trust: A multi-study look at
  the perfect automation schema.
\newblock {\em Theoretical Issues in Ergonomics Science}, 20(4):440--458, 2019.

\bibitem{mayer1995integrative}
R.~C. Mayer, J.~H. Davis, and F.~D. Schoorman.
\newblock An integrative model of organizational trust.
\newblock {\em Academy of management review}, 20(3):709--734, 1995.

\bibitem{MHSW20}
E.~Mayr, N.~Hynek, S.~Salisu, and F.~Windhager.
\newblock Trust in information visualization.
\newblock {\em EuroVis Workshop on Trustworthy Visualization (TrustVis)}, 2019.

\bibitem{mayr2019trust}
E.~Mayr, N.~Hynek, S.~Salisu, and F.~Windhager.
\newblock Trust in information visualization.
\newblock In {\em TrustVis@ EuroVis}, pp. 25--29, 2019.

\bibitem{mcevily2011measuring}
B.~McEvily and M.~Tortoriello.
\newblock Measuring trust in organisational research: Review and
  recommendations.
\newblock {\em Journal of Trust Research}, 1(1):23--63, 2011.

\bibitem{morgan1994commitment}
R.~M. Morgan and S.~D. Hunt.
\newblock The commitment-trust theory of relationship marketing.
\newblock {\em Journal of marketing}, 58(3):20--38, 1994.

\bibitem{mosca2021does}
A.~Mosca, A.~Ottley, and R.~Chang.
\newblock Does interaction improve bayesian reasoning with visualization?
\newblock In {\em Proceedings of the 2021 CHI Conference on Human Factors in
  Computing Systems}, pp. 1--14, 2021.

\bibitem{ottley2012visually}
A.~Ottley, B.~Metevier, and R.~Chang.
\newblock Visually communicating bayesian statistics to laypersons.
\newblock 2012.

\bibitem{padilla2022multiple}
L.~Padilla, R.~Fygenson, S.~C. Castro, and E.~Bertini.
\newblock Multiple forecast visualizations (mfvs): Trade-offs in trust and
  performance in multiple covid-19 forecast visualizations.
\newblock {\em IEEE Transactions on Visualization and Computer Graphics},
  29(1):12--22, 2022.

\bibitem{pandey2023mini}
S.~Pandey and A.~Ottley.
\newblock Mini-vlat: A short and effective measure of visualization literacy.
\newblock In {\em Computer Graphics Forum}, vol.~42, 2023.

\bibitem{pavlou2003consumer}
P.~A. Pavlou.
\newblock Consumer acceptance of electronic commerce: Integrating trust and
  risk with the technology acceptance model.
\newblock {\em International journal of electronic commerce}, 7(3):101--134,
  2003.

\bibitem{ICT}
L.~P. Robert, A.~R. Denis, and Y.-T.~C. Hung.
\newblock Individual swift trust and knowledge-based trust in face-to-face and
  virtual team members.
\newblock {\em Journal of Management Information Systems}, 26(2):241--279,
  2009.

\bibitem{rossi2018impact}
A.~Rossi, K.~Dautenhahn, K.~L. Koay, and M.~L. Walters.
\newblock The impact of peoples’ personal dispositions and personalities on
  their trust of robots in an emergency scenario.
\newblock {\em Paladyn, Journal of Behavioral Robotics}, 9(1):137--154, 2018.

\bibitem{rotter1967new}
J.~B. Rotter.
\newblock A new scale for the measurement of interpersonal trust.
\newblock {\em Journal of personality}, 1967.

\bibitem{rotter1971generalized}
J.~B. Rotter.
\newblock Generalized expectancies for interpersonal trust.
\newblock {\em American psychologist}, 26(5):443, 1971.

\bibitem{rotter1980interpersonal}
J.~B. Rotter.
\newblock Interpersonal trust, trustworthiness, and gullibility.
\newblock {\em American psychologist}, 35(1):1, 1980.

\bibitem{rousseau1998not}
D.~M. Rousseau, S.~B. Sitkin, R.~S. Burt, and C.~Camerer.
\newblock Not so different after all: A cross-discipline view of trust.
\newblock {\em Academy of management review}, 23(3):393--404, 1998.

\bibitem{sabel1993studied}
C.~F. Sabel.
\newblock Studied trust: Building new forms of cooperation in a volatile
  economy.
\newblock {\em Human relations}, 46(9):1133--1170, 1993.

\bibitem{sacha2015role}
D.~Sacha, H.~Senaratne, B.~C. Kwon, G.~Ellis, and D.~A. Keim.
\newblock The role of uncertainty, awareness, and trust in visual analytics.
\newblock {\em IEEE transactions on visualization and computer graphics},
  22(1):240--249, 2015.

\bibitem{schaefer2016measuring}
K.~E. Schaefer.
\newblock Measuring trust in human robot interactions: Development of the
  “trust perception scale-hri”.
\newblock In {\em Robust intelligence and trust in autonomous systems}, pp.
  191--218. Springer, 2016.

\bibitem{ML}
P.~Schmidt and F.~Biessmann.
\newblock Quantifying interpretability and trust in machine learning systems.
\newblock {\em arXiv:1901.08558 [cs, stat]}, 2019.

\bibitem{shu2018human}
P.~Shu, C.~Min, I.~Bodala, S.~Nikolaidis, D.~Hsu, and H.~Soh.
\newblock Human trust in robot capabilities across tasks.
\newblock In {\em Companion of the 2018 ACM/IEEE international conference on
  human-robot interaction}, pp. 241--242, 2018.

\bibitem{simmel2004philosophy}
G.~Simmel.
\newblock {\em The philosophy of money}.
\newblock Psychology Press, 2004.

\bibitem{ullman2018does}
D.~Ullman and B.~F. Malle.
\newblock What does it mean to trust a robot? steps toward a multidimensional
  measure of trust.
\newblock In {\em Companion of the 2018 acm/ieee international conference on
  human-robot interaction}, pp. 263--264, 2018.

\bibitem{xiong2019examining}
C.~Xiong, L.~Padilla, K.~Grayson, and S.~Franconeri.
\newblock Examining the components of trust in map-based visualizations.
\newblock In {\em 1st EuroVis Workshop on Trustworthy Visualization, TrustVis},
  pp. 19--23, 2019.

\bibitem{zhou2019effects}
J.~Zhou, Z.~Li, H.~Hu, K.~Yu, F.~Chen, Z.~Li, and Y.~Wang.
\newblock Effects of influence on user trust in predictive decision making.
\newblock In {\em Extended Abstracts of the 2019 CHI Conference on Human
  Factors in Computing Systems}, pp. 1--6, 2019.

\end{thebibliography}
\end{document}